\documentclass[cits,hyper]{JINST}
\usepackage{ucs}
\usepackage{url}
\usepackage[utf8]{inputenc}
\usepackage{pslatex}
\usepackage{array}
\newcolumntype{L}[1]{>{\raggedright\let\newline\\\arraybackslash\hspace{0pt}}m{#1}}
\newcolumntype{C}[1]{>{\centering\let\newline\\\arraybackslash\hspace{0pt}}m{#1}}
\newcolumntype{R}[1]{>{\raggedleft\let\newline\\\arraybackslash\hspace{0pt}}m{#1}}
\usepackage{multirow}
\usepackage{graphicx}
\graphicspath{{./rysunki/}{./rysunki.src/}}
\newcommand{\DescriptorManager}{{\em Descriptor Manager}}
\newcommand{\CmdProcessor}{{\em Command Processor}}
\newcommand{\EthernetSender}{{\em Ethernet Sender}}
\newcommand{\EthernetReceiver}{{\em Ethernet Receiver}}

\newcommand{\AckFIFO}{{\em Ack \& Cmd FIFO}}
\newcommand{\NlogPktsInFPGA}{{N_{FPGA}}}
\newcommand{\NlogPktsInCPU}{{N_{CPU}}}

\newcommand{\HeadPtr}{{\em Head pointer}}
\newcommand{\TailPtr}{{\em Tail pointer}}
\newcommand{\wzcode}[1]{{\em #1}}
\newcommand{\wzhex}[1]{{\tt #1}}
\newcommand{\wzcite}[1]{\cite{#1}}
\newcommand{\wzurl}[1]{\href{#1}{\url{#1}}}
\newcommand{\wzonline}[1]{}

\author{Wojciech M. Zabolotny$^a$\thanks{Corresponding
author.}\\
\llap{$^a$}Institute of Electronic Systems, Warsaw University of Technology\\
  ul. Nowowiejska 15/19, 00-665 Warszawa, Poland\\
  E-mail: \email{wzab@ise.pw.edu.pl}}

\title{Low latency protocol for transmission of measurement data from FPGA
to Linux computer via 10~Gbps Ethernet link}
 \abstract{
  This paper presents FADE-10G -- an integrated solution for modern multichannel measurement systems.
  Its main aim is a low latency, reliable transmission of measurement data from FPGA-based
  front-end electronic boards (FEBs) to a~computer-based node in the~Data Acquisition System (DAQ),
  using a~standard Ethernet 1~Gbps or 10~Gbps link.
  In addition to transmission of data, the~system allows the~user to send reliably simple control
  commands from DAQ to FEB and to receive responses.

  The~aim of the~work is to provide a~possible simple base solution, which can be adapted
  by the~end user to his or her particular needs.
  Therefore, the emphasis is put on the~minimal consumption of FPGA resources in FEB
  and the~minimal CPU load in the~DAQ computer.
  The~open source implementation of the~FPGA IP core and the~Linux kernel driver published 
  under permissive license facilitates modifications and reuse of the~solution.

  The~system has been successfully tested in real hardware, both with 1~Gbps and 10~Gbps links.
  
 }
\begin{document}
 \section{Introduction}
 In modern multichannel measurement systems, it is often necessary to transfer multiple data
 streams from detectors to computers responsible for processing of data.
 Especially the~introduction of a~triggerless approach in High Energy Physics (HEP) experiments
 (\wzcite{trigger-less-PANDA, trigger-less-LHCb, trigger-less-FLES, trigger-less-Mu3E})
 increased demand on the~amount of data that must be transferred to the~Data Acquisition (DAQ) System,
 and therefore also on the~number of links that must be provided.
 The~Front End Boards (FEB) are typically built using FPGA chips, which nowadays are 
 often equipped with gigabit or multi-gigabit transceivers. That enables the~implementation of a~broad range of
 high-speed communication interfaces~\wzcite{url-altera-transceiver-portfolio,url-xilinx-high-speed-serial}.
 When selecting the~appropriate solution, we must take into account additional requirements like a~length
 of the~link (which in some experiments may reach even a~few hundred meters~\wzcite{url-cbm-readout}) and an electrical insulation (so optical fiber is preferred).
 To reduce the~total cost of implementation of multiple links, we should use a~standard interface
 to benefit from price reduction thanks to the mass production of transceivers and other components of the~link infrastructure.

 Considering the~requirements mentioned above, it~seems that the~Ethernet link, using the~SFP or SFP+ 
 optical transceivers is the~optimal solution.
 Broad use of the Ethernet technology has resulted in a~significant reduction in the price of components (namely
 SFP+ transceivers)  needed to implement the~10~Gbps Ethernet link both on the~computer side and on the~FPGA side.
 The~achievable price of an optical 10~Gbps SFP+ transceiver is approximately \$85 for a single channel.
 It can be further reduced when ordering a~bigger batch, or when using four channel QSFP+ transceivers
 (price approximately \$280 for four channels).
 \subsection{Required functionality of the~transmission system}
 The~aim of this~work is to create a~minimal but extensible solution. Therefore, it is important
 to define the requirements that should be fulfilled by such a~system.
 \begin{itemize}
 \item Possibility to work with 1~Gbps (for price-sensitive applications) and 10~Gbps (for typical applications) Ethernet links
 \item Reliable transport of data stream with maximal throughput and minimal latency (because the~latency
   directly affects the~amount of memory needed to buffer transmitted but not yet confirmed data).
 \item Possibility to control FEBs and to check their status from the~DAQ side of the~link (even though in a typical DAQ system
   there is yet another separate communication channel for configuration and diagnostics of FEBs).
  \item Open source implementation, that may be modified to suit the~needs of the~particular experiment.
  \item Ability to work with different PHY interfaces (copper or optical), depending on the~needs of the~particular experiment.
 \end{itemize}
 
 \section{State of the~art}
The~standard solution for the~reliable transfer of data via an Ethernet network is the~TCP/IP protocol.
Unfortunately, this protocol has serious disadvantages when used in an FPGA.
It has been optimized mainly for the~transport of data in wide area networks
with multiple routers between communicating devices. Therefore, it contains many features
related to routing of data packets, fragmenting the~packets and with sharing
the~link bandwidth between multiple
connections. The~TCP/IP also assumes that data may be transported via untrusted networks, 
and therefore it implements sophisticated algorithms protecting the~communication against malicious
activity. The~final result is that implementation of the~full TCP/IP stack 
in an FPGA is complex and resource hungry. Some implementations rely on a~CPU implemented
in an FPGA or embedded in an FPGA~\wzcite{url-xapp1026}, but such solutions do not allow full utilization
of the~10~Gbps link throughput.
There are some commercial implementations of 10~Gbps hardware TCP/IP stacks for an FPGA, but they are closed
and expensive solutions~\wzcite{url-accelize-tcpip, url-dgway-toe10g}.

\subsection{Reduced TCP/IP developed at CERN}
An interesting attempt to reduce resource consumption of the~hardware-based TCP/IP implementation
is a~solution developed at CERN and described in~\wzcite{Bauer2014, 1748-0221-8-12-C12039}.
The~authors have reduced the functionality of the~TCP/IP so that it is possible to implement it
in an~FPGA without implementing any soft CPU core.
The~implementation provides only unidirectional transmission. 
The~authors did not implement timestamps, selective acknowledgments or out of band data.
In addition, certain mechanisms have been significantly simplified -- e.g. the~congestion management.
This solution however still relies on the~external DDR memory used as a~TCP socket buffer.
The advantage of this solution is that the~receiver may be a~standard computer with TCP/IP stack provided
by the~operating system. However, this solution also leads to significant CPU load. 
As the~authors state themselves ``Running one 10~Gbps TCP stream can easily saturate one
of the~CPU cores.''

Another significant disadvantage is the~closed source nature of this solution. No sources have been 
released, so it can not be a~basis for an open, extensible solution.
\subsection{Avoiding the~TCP/IP complexity}
To avoid the implementation of a complex TCP/IP stack in the~FPGA and to reduce the~load of CPU
in the~receiving computer, it is desirable to use a~simpler protocol.
Usage of the~UDP protocol instead of TCP is not optimal. The~UDP protocol does not assure
reliable transfer of data, so it is necessary to implement additional mechanisms ensuring reliability.
At the~same time, the~UDP protocol and all IP protocols still require 
significant overhead associated with the~routing of packets (datagrams).
\begin{figure}[t]
   \begin{center}
   \begin{tabular}{c}
   \includegraphics[width=0.7\linewidth]{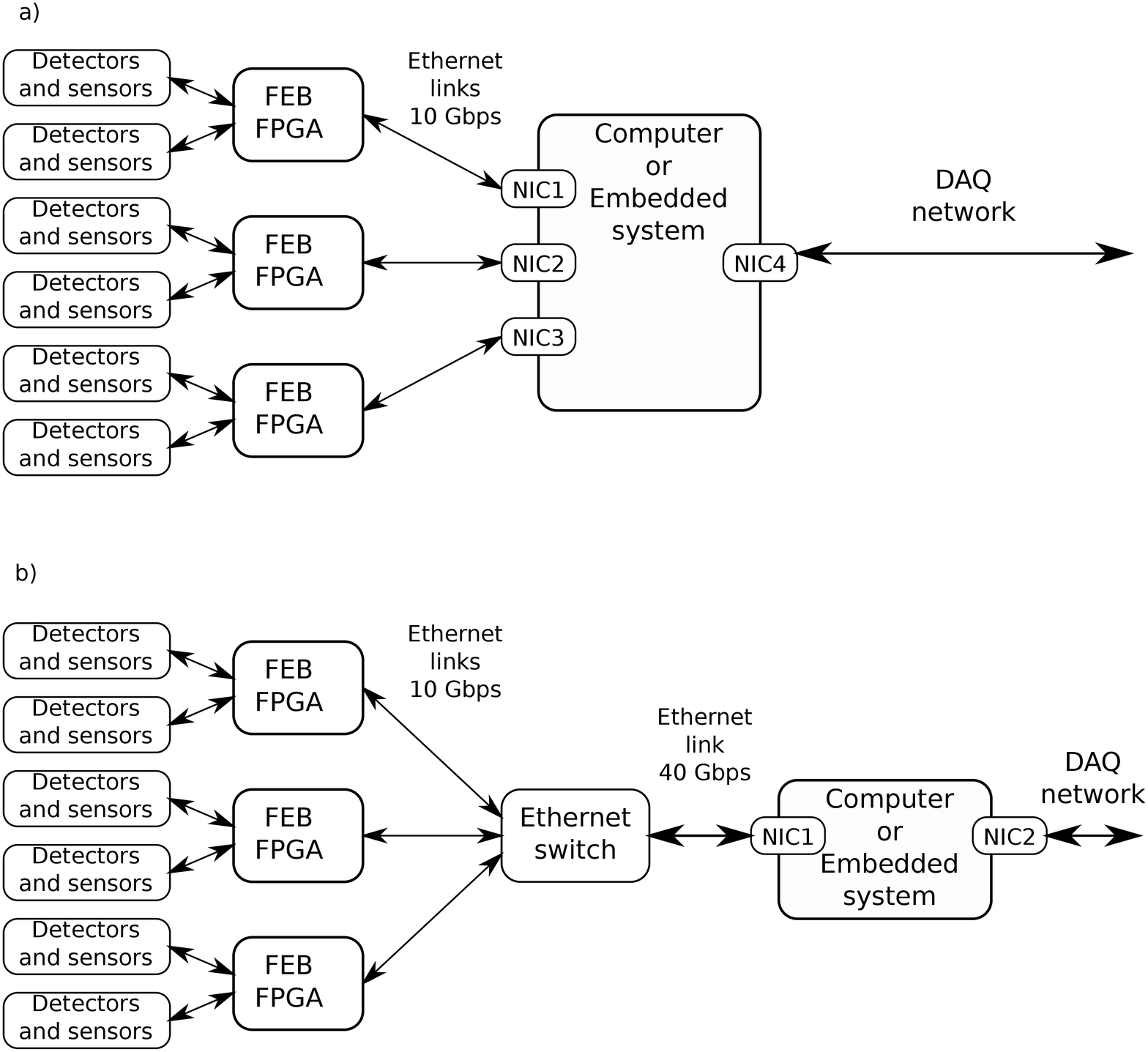}
   \end{tabular}
   \end{center}
   \caption
   { \label{fig:ether-based-daq}
    Possible topologies of Ethernet-based data transmission from FEB to DAQ.
    a) The~case where Ethernet interface speeds in FEBs and the~computer are equal.
    b) The~case where the~computer offers a~faster Ethernet interface.
   }
\end{figure}
However, the~connection between FEBs and DAQ should not contain any routers, as they increase 
link latency, which in turn leads to an increase of memory needed to buffer the~transmitted
and not yet confirmed data.

There are two possible link topologies. In the~case where Ethernet interfaces in both -- FEBs and DAQ
computers have the~same speed the~point-to-point connections will be used
(see Figure~\ref{fig:ether-based-daq}a).
If the~Ethernet interface in the~DAQ computer offers a~higher speed (e.g. 40~Gbps),
it is possible to connect a~few FEBs to a~single network card via a~10~Gbps/40~Gbps switch
(see Figure~\ref{fig:ether-based-daq}b).
For such very simple networks, where Ethernet frames are passed either directly
or via a~Layer 2 network switch, the~best solution is to develop an optimized
Layer 3 protocol using raw Ethernet frames. 
\subsection{Ethernet Proxy -- EPRO as possible solution}
The~protocol and Linux kernel driver based on the~above assumptions was
developed at the~AGH University of Science and Technology and 
described in~\wzcite{mindur-ether-art, mindur-eppro}.
The~proposed solution implements not only a~reliable transport of the~data stream, 
but also some additional functions. 
Those functions include different types and priorities of data,
or the~possibility to send the~same data to more than one destination.
The~protocol is implemented for a~1~Gbps link and uses the~standard Xilinx MAC implementation.
Unfortunately, this solution like the~previous one is not open. The~authors did not publish sources,
so it is not possible to modify it to work with higher speed 10~Gbps links or to adjust it to the
particular experiment's requirements.

\subsection{First version of the~FADE protocol}
Another possible solution is the~author's open source FADE protocol described in~\wzcite{wzab-fade1}.
This protocol provides reliable transmission of data from
an~FPGA to a~computer through 1~Gbps Ethernet links.
The~resource consumption in the~FPGA is kept to a~minimal level and may be adjusted using
a~parametrized VHDL code. Instead of a complex standard MAC, simplified state machines 
are used to receive and send packets. These are sufficient for full-duplex Ethernet links with
granted link bandwidth.
The~initial version of the~FADE protocol worked correctly with 1~Gbps links, but
an attempt to simply modify the~FPGA IP core for operation with 10~Gbps Ethernet PHY
revealed problems with efficiency. 
Therefore, the~whole code was significantly modified.
Modifications included simplification of the~packets management. For example, the~concept of
``sets of packets'' described in~\wzcite{wzab-fade1} was dropped in favor of a~simple
description of the~data stream as a~continuous sequence of packets.
Another modification was the~addition of the~possibility to perform simple control and diagnostic
operations via the~Ethernet link while the~original FADE protocol allowed 
only to send START and STOP commands.

This article describes the~implementation of the~new version of the~FADE protocol named FADE-10G.

\section{Implementation of the~FADE-10G protocol}
The~FADE-10G protocol is aimed at the~transmission of the~continuous data stream
consisting of 64-bit words. To better utilize the~link bandwidth, data are transmitted
using Ethernet jumbo frames. The data packets contain 1024 data words (8192 bytes) and
some additional information (MTU should be set to 9000 in the~network interface configuration).
The~number of data words in a~packet equal to the~power of two
was chosen to simplify packet management both in the~FPGA and in the~receiving computer,
as it is described later.
When the~transmission is stopped, the~last packet may contain fewer data words.
In such a~case, the~last data word contains the~number of valid words in that packet (between 0
and 1023).
Because the~protocol is supposed to be used as the~only protocol in private networks,
the~private, unofficial Ethertype \wzhex{0xfade} is used.
To differentiate frames of the~FADE-10G from the~old FADE frames, and to allow further 
modifications of the~protocol, the~protocol version number is transmitted after the~Ethertype
field. This number is equal to 0x0100 in the~current version%
\footnote{In the~first version of the~FADE protocol, this field contained the~type of the~frame
and could be a~value from the~range 0x01 to 0x05 or the~value 0xa5a5.}.
Because the~Ethernet link does not warrant the reliable delivery of frames,
it is necessary to implement a~simple acknowledgment/retransmission algorithm, which uses
special shorter acknowledgment frames.
Still others short frames are necessary to allow transmission of simple control or diagnostic
commands via Ethernet link. 
The~general structure of the~FADE-10G Ethernet frame is shown in Table~\ref{tab:fade-frame-gen-struc},
and the~payload contained in frames of different types is shown in Table~\ref{tab:fade-frame-payload}.
\begin{table}[tp]
  \caption[General structure of frame]
  { \label{tab:fade-frame-gen-struc}
    Structure of the~Ethernet frames used by the~FADE-10G transmission protocol.
  }
{\small
\begin{tabular}{|C{1.3cm}|C{1.3cm}|C{1.3cm}|C{1.3cm}|C{3cm}|C{3cm}|C{1.3cm}|}
\hline
\multicolumn{3}{|c|}{Standard Ethernet header} & Protocol version & Payload  & Filler & Checksum \\
\hline
Source MAC & Destination MAC & 0xFADE & 0x0100 & Payload bytes & N*0xa5 & FCS \\
\hline
6 bytes & 6 bytes & 2 bytes & 2 bytes & length depends on the~type of frame & variable length used when the frame is too short & 4 bytes \\
\hline
\end{tabular}
}
\end{table}

\begin{table}[tp]
  \caption[Payload in different frames]
  { \label{tab:fade-frame-payload}
    Structure of the~payload in different Ethernet frames used by the~FADE-10G transmission protocol.
  }
{\small
\begin{tabular}{l}
a) Data acknowledgment frame (from computer to FPGA) \\
\begin{tabular}{|C{2cm}|c|c|c|}
  \hline
  0x0003 (ACK) or 0x0004 (NACK) & Frame sequence number & Packet number in the~data stream & Transmission delay \\
  \hline
  2 bytes                       &      2 bytes & 4 bytes       & 4 bytes \\
  \hline
\end{tabular}\\
~\\
b) User command request (from computer to FPGA) \\
\begin{tabular}{|C{2cm}|c|c|c|}
  \hline
  Command code & Command sequence number & Command argument \\
  \hline
  2 bytes      & 2 bytes & 4 bytes\\
  \hline
\end{tabular}\\
~\\
c) Standard data packet (from FPGA to computer)\\
\begin{tabular}{|C{1.5cm}|C{1.5cm}|C{1.5cm}|C{1.5cm}|C{1.5cm}|C{2cm}|}
  \hline
  0xA5A5 & Frame sequence number & Packet number in data stream & Transmission delay & Command response$^\textrm{*}$ & data \\
  \hline
  2 bytes& 2 bytes & 4 bytes & 4 bytes & 12 bytes & 8192 bytes \\
  \hline
\end{tabular}\\
~\\
d) Last data packet (from FPGA to computer)\\
\begin{tabular}{|C{1.5cm}|C{1.5cm}|C{1.5cm}|C{1.5cm}|C{1.5cm}|C{2cm}|C{2cm}|C{2cm}|}
  \hline
  0xA5A6 & Frame sequence number & Packet number in data stream & Transmission delay & Command response$^\textrm{*}$ & data & number of valid words \\
  \hline
  2 bytes& 2 bytes & 4 bytes & 4 bytes & 12 bytes & 8184 bytes, not all must be valid & 8 bytes \\
  \hline
\end{tabular}\\
~\\
e) Command response packet (from FPGA to computer)\\
\begin{tabular}{|C{2cm}|C{2cm}|}
  \hline
  Filler & Command response$^\textrm{*}$\\
  \hline
  2 bytes& 12 bytes\\
  \hline
\end{tabular}\\
~\\
 
$^\textrm{*}$ Command response field in the~data packet or command response packet:\\
\begin{tabular}{|C{2cm}|C{2cm}|C{3cm}|}
  \hline
  Command code & Command sequence number & User defined return value\\
  \hline
  2 bytes& 2 bytes & 8 bytes\\
  \hline
\end{tabular}\\
\end{tabular}
}
\end{table}

\subsection{Implementation of the~protocol in the~FPGA}
\label{sec:impl-in-fpga}
The reliable transmission of data via an unreliable channel (like an~Ethernet link) requires
retransmission. Therefore, it is necessary to buffer the~data that have been transmitted,
but have not yet been confirmed by the~receiving computer.

To keep the~algorithm controlling the~retransmission as simple as possible,
the~memory buffer in the~FPGA has a~length of $M=2^\NlogPktsInFPGA$ data packets.
Each data packet is 1024 words (8192 bytes) long. Thus, the~lower bits of the~number of the~data packet in the~data stream may be directly used to define its position
in the~memory buffer. The~length of this buffer also defines the~\wzcode{transmission window}
of the~protocol. At every moment, only a~packet from a~certain set of $M$ consecutive packets
may be transmitted via the~link. The~$\NlogPktsInFPGA$ value may be configured before synthesis of the~core
and compilation of the~protocol driver (described in Section~\ref{sec:kernel-driver}).

\begin{figure}[t]
   \begin{center}
     \begin{tabular}{c}
       \includegraphics[width=0.7\linewidth]{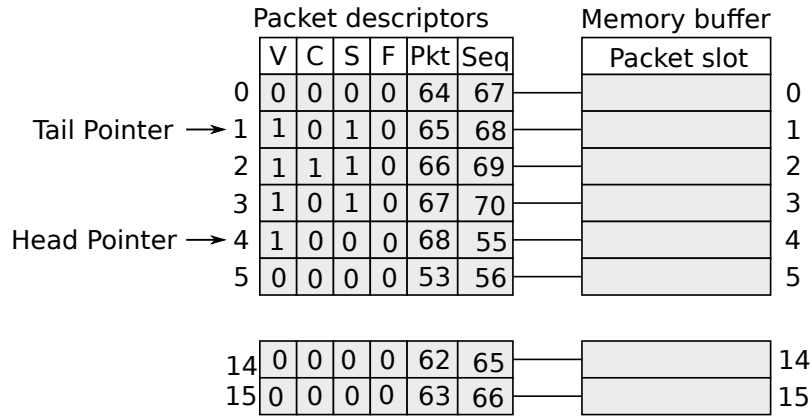}
     \end{tabular}
   \end{center}
   \caption
       { \label{fig:packet-descriptors}
         Data packets in the~FPGA memory and their descriptors shortly after the start of transmission.
         Bit flags: V-Valid, S-Sent, C-Confirmed,
         F-Flushed (used when the transmission is finished). The~``Pkt'' field stores the~31-bit number of the packet
         in the~data stream. The~"Seq" field stores the~16-bit frame sequence number used by the~fast
         retransmission algorithm.
         The~packets associated with descriptors 1 and 3 contain valid data. They have been sent
         but are not confirmed yet. Please note, that the sequence numbers are higher than the packet numbers because
         three packets were retransmitted before.
         The~packet associated with descriptor 2 contains valid data, it has been sent and is confirmed.
         The~packet associated with descriptor 4 contains valid data, but it has not yet been sent (therefore
         its sequence number is lower, as it is the~sequence number of the~packet
         that previously occupied this slot).
         Other descriptors are free. Therefore, their flags are cleared.
       }
\end{figure}

Each packet is associated with its descriptor shown in Figure~\ref{fig:packet-descriptors}.

\begin{figure}[t]
   \begin{center}
   \begin{tabular}{c}
   \includegraphics[width=0.9\linewidth]{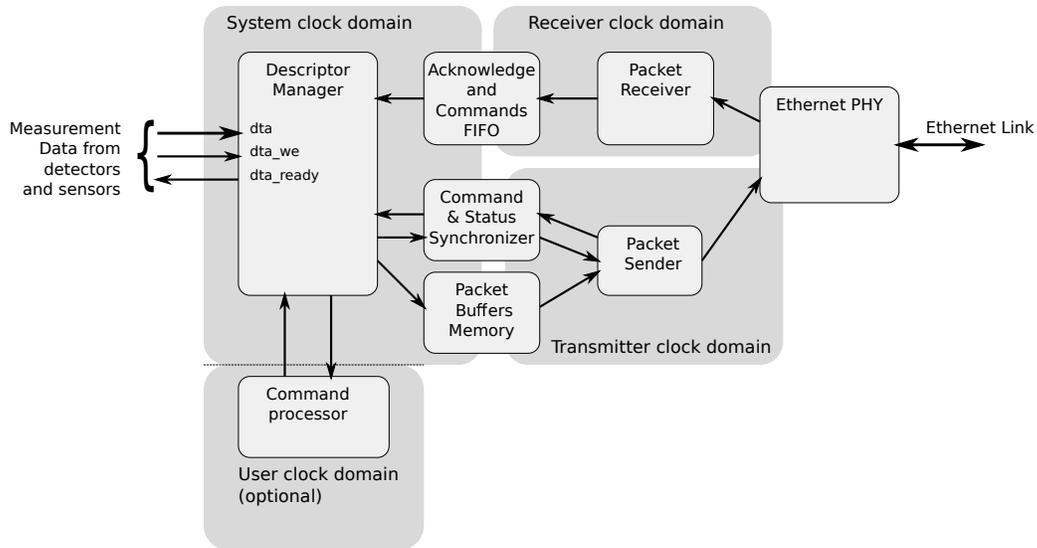}
   \end{tabular}
   \end{center}
   \caption
   { \label{fig:ip-core-structure}
    Structure of the~FPGA IP core supporting the~FADE-10G protocol.
   }
\end{figure}

The~structure of the~IP core implemented in the~FPGA is shown in Figure~\ref{fig:ip-core-structure}.
The~\EthernetReceiver{} and \EthernetSender{} blocks are simple state machines, replacing the~standard 
Ethernet MAC. They are connected to the~external Ethernet PHY or the~internal
Ethernet PHY equivalent implemented
in the~FPGA -- like the~Xilinx PCS/PMA core~\wzcite{url-xilinx-10g-pcs-pma}.

64-bit data words provided by the~data source are written to the~data packet 
pointed by the~\HeadPtr{}. When this packet is filled, it is marked
as ready for transmission (V=1). Then the~\DescriptorManager{} checks if it is
possible to move the~\HeadPtr{} to the~next position. If the~next position is
the~one pointed by the~\TailPtr{}, it means that the~buffer is full. In this case,
the~ready status of the~core is deasserted until the~packet pointed by the~\TailPtr{}
 is acknowledged, and the~\TailPtr{} is moved to the~next position.

The~\EthernetReceiver{} block receives packets, checks their checksum, and writes
information from correctly received  packages to the~{\em Acknowledgment and Commands FIFO} 
(\AckFIFO{}).
Additionally, the~\EthernetReceiver{} itself executes a~few high priority
commands like START, STOP, and RESET. The~START and STOP commands are still written
to the~\AckFIFO{} to ensure generation of their confirmation.
The~RESET command causes the reset of the~whole FADE-10G core  and is therefore not confirmed at all.

The~\DescriptorManager{} reads commands from the~\AckFIFO{}. If the~received command is 
the~packet acknowledgment (ACK) or negative packet acknowledgment (NACK), the~\DescriptorManager{} handles 
it itself, as these commands are not confirmed. Other commands are passed to 
the~\CmdProcessor{}, which executes the~command and generates their confirmation.

The~packet acknowledgment (ACK) command sets the~C (Confirmed) flag in the~descriptor
of the~acknowledged packet if this packet is still kept in the~buffer%
\footnote{It is possible that the~core receives a~delayed duplicated
 acknowledgment packet. In that case, the~buffer
 no longer  contains the~corresponding descriptor.}.
If the~received ACK packet contains a~packet number bigger than the~number 
of the~last transmitted packet%
\footnote{The~packet number wraps every $2^{32}$ packets. Therefore comparison of those
 numbers is defined as follows:\\ $N_1 > N_2$ if $(N_1 - N_2)\pmod{2^{32}} \le {2^{31}}$.}%
, a~protocol error is detected.

After all commands available from the~\AckFIFO{} are executed the~\EthernetReceiver{} block tries
to move the~\TailPtr{} freeing all packets that have the~C flag set
in their descriptor. All flags in descriptors of freed buffers are cleared.
After that operation, if there is a~free place in the~buffer, the~ready status of the~core
is asserted again.
  
Another activity performed by the~\DescriptorManager{} is the~transmission and retransmission
of packets. It continuously browses the~packet buffer and finds packets that have
 the~V flag set, but the~C flag unset. Those packets are passed to 
the~\EthernetSender{} block for transmission or retransmission.
  
The~last hardware block is the~\CmdProcessor{}, which may work in the~same clock domain
as the~\DescriptorManager{} but may also operate in another (even its own) clock domain.
The~\CmdProcessor{} executes the~received command and after the~result or status is ready
it builds the~command response and passes it to the~\DescriptorManager{}. The~command
response is then transmitted either in the~nearest data packet or the~dedicated
command response packet (if no data packet is currently waiting for transmission
or retransmission).
  
The~core counts transmitted data packets and retransmitted data packets 
to avoid network congestion or computer overload.
The~ratio of those counts is then calculated.
If the detected ratio of retransmitted packets is too high which may be a~symptom of an overload,
the~delay between transmitted packets is increased.
If the~ratio of retransmitted packets is very small, this delay is decreased.
Thresholds used by the~delay adaptation algorithm are parametrized and may be changed
before synthesis of the~core. For debugging purposes, current transmission delay is
included in packets sent from the~FPGA to the~computer (field 
\wzcode{Transmission Delay} in Table~{\ref{tab:fade-frame-payload}}).

\subsection{Early retransmission mechanism}
\label{sec:early-retransm}%
The~retransmission algorithm described has one significant disadvantage. If the~packet pointed by
the~\TailPtr{} (or its acknowledgment) is lost, the~space in the~packet buffer will not be freed
until this packet is retransmitted again and successfully confirmed. In the~described implementation,
 this packet will be only retransmitted after all other pending packets
are transmitted or retransmitted. Therefore, the~core will not accept new data for a~significant 
amount of time.

The~performance of the~algorithm may be improved,  if such a~packet is retransmitted as soon
 as its loss (or loss of its acknowledgment) is detected.
The~clear sign of such an event is when the~core receives acknowledgment of the~packet
that has been transmitted after that one.
Such a~solution is similar to the~``Fast retransmit'' used in the~TCP protocol~\wzcite{rfc2581}.

In the~simplest solution after reception of the~acknowledgment of any packet all unconfirmed packets with the packet number smaller than the one received will be retransmitted.
\begin{figure}[t]
   \begin{center}
   \begin{tabular}{c}
   \includegraphics[width=0.7\linewidth]{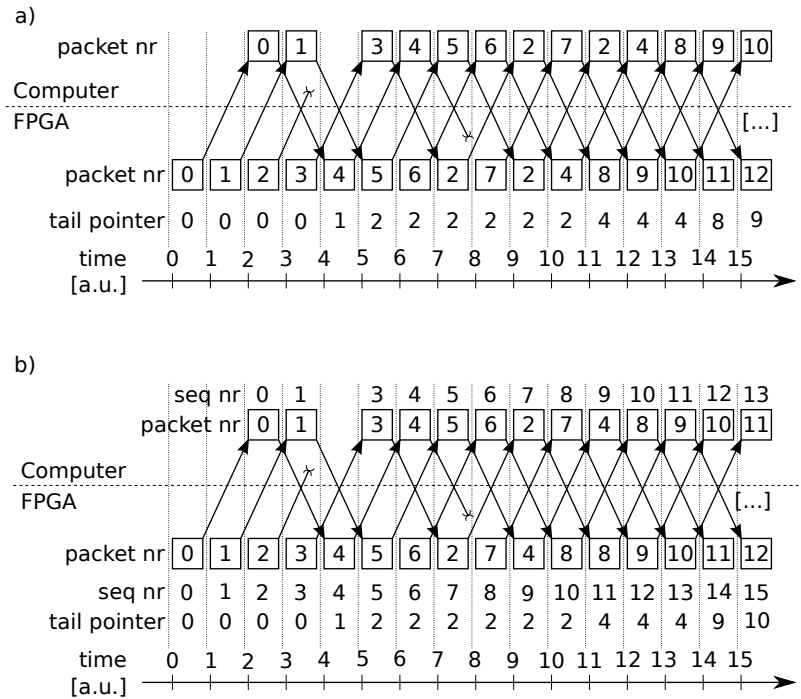}
   \end{tabular}
   \end{center}
   \caption
   { \label{fig:early-retransm}
    Operation of the~early retransmission mechanism: a) without sequence numbers, b) with sequence numbers.
    In both cases, packet 2 is lost, and the~ACK for packet 4 is lost. At time=7, packet 3 gets confirmed,
    but no ACK for packet 2 has been received earlier. Therefore, packet 2 is scheduled for immediate retransmission.
    At time=9, packet 5 gets confirmed, but neither packet 2 nor packet 4 have been confirmed yet. Therefore in case
    (a) both those packets are scheduled for immediate retransmission. In case (b), the~sequence number is checked. At that
    moment, the~last sequence number for packet 2 is equal to 7 and for packet 4 to 4. As the~received ACK for packet 5 
    has sequence number equal to 5, only packet 4 is scheduled for immediate retransmission.
   }
\end{figure}
Unfortunately, this kind of simplistic implementation based only on the~number of the~packet in
the~data stream is not optimal. If loss of yet another packet is detected before the ``early retransmitted'' packet is confirmed this packet will be unnecessarily retransmitted
once again (see Figure~\ref{fig:early-retransm}a). To prevent this, the~data packets are labeled additionally with 
the~\wzcode{frame sequence number} incremented after every transmission.
The~last \wzcode{frame sequence number} used to transmit the~particular data packet is stored
in the~packet's descriptor (the~``Seq'' field in Figure~\ref{fig:packet-descriptors}).
This \wzcode{frame sequence number} is copied to the~acknowledgment packet.
When loss of the~packet is detected, it is possible to retransmit early only those 
packets that have a~last \wzcode{frame sequence number} smaller%
\footnote{The~frame sequence number wraps every $2^{16}$ packets. Therefore comparison of those
 numbers is defined as follows:\\ $N_1 > N_2$ if $(N_1 - N_2)\pmod{2^{16}} \le {2^{15}}$.}
 than the~acknowledgment packet
just received (see Figure~\ref{fig:early-retransm}b).

\subsection{Execution of user commands}
\label{sec:fpga-command-execution}
To ensure that each command is delivered reliably, executed exactly once and its~results
are delivered successfully to the~computer, the~\wzcode{command sequence numbers} (CSNs) are used.

Whenever a~new command is sent to the~FPGA core, the~CSN is increased. That enables discarding of
possible duplicated responses to previous commands. After a new command is sent, the~computer 
waits for the~response for a~certain configurable amount of time.
If the~computer does not receive the~command response packet in the~declared
time period, it states that either the~command packet or the~response packet were lost.
In that case, the~computer resends the~same command once again.

When the~FPGA core correctly receives the~command packet it first checks its CSN.
If it is the~same as in the~last serviced command (which means that the~response packet
was lost, and command was resent), the~core only resends the~response for that last command.
If the~CSN is different, the~core stores it, executes the~command 
and then sends the~response packet with the~same CSN.

The~command response is sent in the~pending data frame if it is available or (when
currently no data packet is waiting for transmission or retransmission)
in the~dedicated command response packet (see Table~\ref{tab:fade-frame-payload}~c--e).

\subsection{Resource consumption}
\label{sec:fpga-synthesis}
\begin{table}[tp]
  \caption[Resource usage]
  { \label{tab:fade-resource-usage}
    The per-link resource usage of the FADE-10G core synthesized
    with different size of the memory buffer
    for different FPGA chips.
  }
{\small
\begin{tabular}{|C{3cm}|C{2cm}|C{2.5cm}|C{2.5cm}|C{2.5cm}|}
\hline FPGA chip and buffer length & Resource & Used units & Available units & The percentage of usage\\ 
\hline
\multirow{4}{*}{\begin{minipage}[c]{3cm}\begin{center}
   Kintex 7 \\ xc7k32tffg900-2 \\ $\NlogPktsInFPGA=32$
   \end{center}
   \end{minipage}}
   & Slices & 801 & 50,950 & 1.57\% \\ 
\cline{2-5} & Slice LUTs & 2,107 & 203,800 & 1.03\% \\ 
\cline{2-5}  & Slice registers & 1,240 & 407,600 & 0.3\% \\ 
\cline{2-5}  & BRAM tiles & 65 & 445 & 14.6\% \\ 
\hline 
\multirow{4}{*}{\begin{minipage}[c]{3cm}\begin{center}
   Kintex 7 \\ xc7k32tffg900-2 \\ $\NlogPktsInFPGA=16$
   \end{center}
   \end{minipage}}
   & Slices & 756 & 50,950 & 1.48\% \\ 
\cline{2-5}  & Slice LUTs & 2,065 & 203,800 & 1.01\% \\ 
\cline{2-5}  & Slice registers & 1234 & 407,600 & 0.3\% \\ 
\cline{2-5}  & BRAM tiles & 33 & 445 & 7.42\% \\ 
\hline
\multirow{4}{*}{\begin{minipage}[c]{3cm}\begin{center}
   Spartan 6\\ xc6slx45csg324-2 \\ $\NlogPktsInFPGA=16$
   \end{center}
   \end{minipage}}
   & Slices & 611 & 6,822 & 8.96\% \\ 
\cline{2-5}  & Slice LUTs & 1599 & 27,288 & 5.86\% \\ 
\cline{2-5}  & Slice registers & 1227 & 54576 & 2.25\% \\ 
\cline{2-5}  & BRAM blocks & 68 & 116 & 58.6\% \\ 
\hline 
\end{tabular} 
}
\end{table}
The~FADE-10G core supporting 10~Gbps links was successfully synthesized for the Kintex 7 xc7k325tffg900-2 FPGA used in KC705~\wzcite{url-kc705} and AFCK~\wzcite{url-afck} boards. 
The version supporting 1~Gbps links was successfully synthesized for the Spartan 6 xc6slx45csg324-2 FPGA used in Atlys boards~\wzcite{url-atlys}.
Synthesis for the Kintex 7 FPGA was performed for two sizes of the memory buffer ($\NlogPktsInFPGA=32$ and $\NlogPktsInFPGA=16$).
Due to the limited amount of internal memory synthesis for the Spartan 6 FPGA was performed only with $\NlogPktsInFPGA=16$.
Results of the synthesis are presented in Table~\ref{tab:fade-resource-usage}.
It is visible that the~FADE-10G core leaves a reasonable amount of logic resources available for the user
to implement FEB blocks. 
For $\NlogPktsInFPGA=32$ the xc7k325tffg900-2 can easily accommodate the~FADE-10G core operating four 10~Gbps links.
For $\NlogPktsInFPGA=16$ the same chip can work with even eight such links.

\section{Linux driver}
 \label{sec:kernel-driver}
 GNU/Linux is widely used in modern data acquisition systems. As a free and open source system,
 it is a~perfect platform for such an open solution as the~one 
 proposed in this paper.
 
 Because FADE-10G uses a~non-standard Ethernet protocol, it is necessary to implement a~dedicated
 kernel driver as a~\wzcode{protocol handler} responsible for the reception of the~Ethernet frames
 of type \wzhex{0xfade}.
 Similarly to the solutions described in~\wzcite{mindur-ether-art} and~\wzcite{wzab-fade1}, 
 the~\wzcode{protocol handler} is installed using the~\wzcode{dev\_add\_pack} function.
 Whenever the~Ethernet frame with \wzhex{0xfade} type is received, the~callback function in
 the~driver is called.

 The~driver may service one or more FPGA-based FEBs. They can be connected to separate
 Ethernet cards, or (via a~switch) to the~same Ethernet card (see Figure~\ref{fig:ether-based-daq}).
 Each connected FPGA-based FEB is serviced via a~dedicated character device (\wzcode{/dev/l3\_fpga\%d},
 where \wzcode{\%d} is replaced with subsequent numbers starting from 0).
 The~maximum number of serviced FEBs is declared when loading the~driver using the~\wzcode{max\_slaves}
 parameter.
 The character device may be opened and configured for communication with the~FPGA (slave)
 using the particular MAC address. After that a~\wzcode{slave context} is created, describing the~state
 of communication with that slave.
 One of the~components of the~\wzcode{slave context} is the~\wzcode{receiver packet buffer},
 which stores received data. The~amount of memory available in the~computer is significantly higher than the~amount 
 of internal memory in the~FPGA. Therefore, this buffer may be much longer than 
 the~memory buffer in the~FPGA core (which
 has a~length of $2^\NlogPktsInFPGA$ packets as described in Section~\ref{sec:impl-in-fpga}).
 Its length is chosen to be $2^\NlogPktsInCPU$ packets. Thus, the~lower bits of the~packet number in the
 data stream may be directly used as the~number of the~corresponding \wzcode{packet slot} in 
 the~\wzcode{receiver packet buffer}.
 That is a circular buffer with the \HeadPtr{} and the \TailPtr{} pointing respectively
 to the~next byte to be written and to the~last byte not yet read.

 \subsection{Packet reception routine}

 The~callback function \wzcode{my\_proto\_rcv} is called when the~packet of \wzcode{0xfade} type is received.
 It first checks if the~packet arrived from the~correct (``opened'') FEB. If not, the~driver
 sends a ``reset'' command to the~misbehaving FEB.
 If the~correct FEB slave is found, further operations are performed on the~\wzcode{slave context}
 of that FEB.

 The~function checks the~protocol version. If it is incorrect, the~appropriate error flag is set, and the~packet is dropped.
 Then the~type of the~received packet is checked. If it is a~command response packet, and if there is a~thread
 waiting for completion of this command, the~result is copied from the~packet to the~user-space buffer.  
Then the~waiting thread is woken up, and the~function returns.
 If the~packet is neither a~response packet nor a~data packet, the~error flag is set, and the~packet is dropped.

 If none of the above conditions applies, the~packet is handled as a~data packet.
 First the~function checks the~command response section. If there is a~thread waiting
 for the~completion of the~corresponding command, the result is copied to the~user space. Then the waiting thread is woken up 
 (like in the~case of a~dedicated command response packet).
 Afterward, the data part of the packet is handled.
 The~\wzcode{receiver packet buffer} stores the~32-bit \wzcode{packet number} of the~last
 received and confirmed packet for each \wzcode{packet slot}.
 The~number of the~received packet in the~data stream (see Table~\ref{tab:fade-frame-payload})
 is compared to the~\wzcode{packet numbers} of packets 
 currently stored in the~\wzcode{receiver packet buffer}.
 If the~packet is already received and confirmed, or if the~packet is ``older''%
 \footnote{The~``age'' of packets is checked by subtracting their numbers modulo $2^{32}$. The result below $2^{31}$
is considered to be a~positive number.} than packets in the~buffer, it is assumed that the~confirmation was possibly lost. In this case, the~function simply marks that the~packet should be confirmed once again. If the~packet is ``newer'' than the~packets in the~current
 \wzcode{transmission window}, it means that a~protocol error has occurred -- the~function sets
 the~appropriate error flag and drops the~packet.
 If none of the~above conditions applies, the~packet contains new, unconfirmed data. 
 The~length of the~packet is verified, and the~function checks if there  is enough free space in the~\wzcode{receiver packet buffer}.
 If not, it drops the~packet (it will be retransmitted again by the~FEB).
 If there is enough free space, data from the~packet is copied to the~corresponding \wzcode{packet slot}.
 
 If the~received packet is the~``last unconfirmed'', the~routine updates the~\HeadPtr.
 After that if the~amount of data available in the~buffer is higher than the~``receiver wake-up threshold'' set by the~user application,
 the~receiving thread is woken up.
If the received packet it the~``last data packet'' (see Table~\ref{tab:fade-frame-payload}d), the~last packet flag is also set.
 If the~last packet flag is set and all packets are confirmed, the~``end of transmission'' flag is set, and
 the~receiving thread is also woken up to receive the last part of data.
 In each case, if required, the~confirmation packet is prepared and scheduled for transmission.

 \subsection{Communication with the~user application}
To avoid conflicts when controlling different slaves, each character device (\wzcode{/dev/l3\_fpga\%d})
may be open only once, by one application.
However, the~user application may perform two different activities: reception of data and sending of control commands.
Commands are serviced in a~synchronous way and the~thread sending the~command is put to sleep
until the~command is executed, and the~response is received.
When the~data is transmitted at high speed, it is unacceptable to stop the~data reception until the
command is executed. Therefore, the~user application should start an additional thread after
 opening the~device so that the~reception and processing of data and the~execution
 of control commands are handled in separate threads.

To avoid overhead associated with copying data, the~\wzcode{receiver packet buffer} for each slave
should be mapped into the~appropriate application's memory using the~driver's \wzcode{mmap} function.
Therefore, the~data is copied only once from the~socket buffer delivered 
by the~Network Interface Card (NIC) driver
to the~shared kernel \wzcode{receiver packet buffer}%
\footnote{There are technologies offering the~true zero copy handling of network data 
like ``Direct NIC Access''~\wzcite{url-direct-nic-access} or ``PF\_RING ZC''~\wzcite{url-pf-ring-zc}. However it is not
clear whether they can be used to create a~continuous representation of the~received data in the~user application
memory without additional copying. The~PF\_RING ZC still requires single copying of the~data when used with
a~standard NIC.}.
Of course, the~access to such shared memory must be appropriately synchronized.
That is achieved using the~\wzcode{ioctl} function. The~driver implements a~set of ioctl commands summarized
in Table~\ref{tab:ioctls}.
\begin{table}[tp]
  \caption[example]
  { \label{tab:ioctls}
    The~\wzcode{ioctl} commands implemented in the~kernel module, to support communication
    with the~user-space application.
  }
\begin{center}
{\small
\begin{tabular}{|c|p{10cm}|}
\hline
IOCTL code & Description of the~commands\\
\hline
L3\_V1\_IOC\_SETWAKEUP & Sets the~amount of data bytes that must be available in the~circular buffer before
   the~user-space application is woken up.\\
L3\_V1\_IOC\_GETBUFLEN  & Returns the~length of the~circular buffer associated with a~particular FEB.\\
L3\_V1\_IOC\_READPTRS & Returns the~number of available data bytes and the~positions of the~\HeadPtr{} and \TailPtr{}
   in the~circular buffer associated with a~particular FEB. Provides necessary synchronization when accessing the~pointers.\\
L3\_V1\_IOC\_WRITEPTRS & Should be called with the~number of bytes processed by the~application.
   Provides necessary synchronization and updates
   the~\TailPtr{} in the~circular buffer associated with a~particular FEB.\\
L3\_V1\_IOC\_GETMAC & Associates the~FEB identified by the~given MAC and connected to the~given network interface
   with the~particular character device.\\
L3\_V1\_IOC\_STARTMAC & Starts the~transmission from the~previously associated FEB.\\
L3\_V1\_IOC\_STOPMAC  & Stops the~transmission from the~FEB associated with the~particular character device.\\
L3\_V1\_IOC\_FREEMAC & Disassociates FEB from the~particular character device.\\
L3\_V1\_IOC\_RESETMAC & Resets the~FADE-10G core in the~FEB associated with a~particular character device.\\
L3\_V1\_IOC\_USERCMD & Sends the~user command to the~FEB associated with a~particular character device. This command
   puts the~current thread to sleep until the~command is executed, and the~result is sent back.\\
\hline
\end{tabular}
}
\end{center}
\end{table}
\subsubsection{Reception of data}
The~user application may read the current positions of the~\HeadPtr{} and \TailPtr{} in its receiver packets buffer using the
\wzcode{L3\_V1\_IOC\_READPTRS} ioctl command. This command ensures appropriate synchronization
so that the~stable \HeadPtr{} values are read. Additionally, this command returns the~number of available bytes in the~receiver packet buffer.
To avoid active waiting for data, the~application may define (with the~\wzcode{L3\_V1\_IOC\_SETWAKEUP} ioctl command)
how many bytes of data must be available in the~receiver packet buffer before the~receiver thread is woken up.
The~thread will be woken up also when the~transmission finishes, even if the~number
of available bytes is below the~defined threshold.
\subsubsection{Sending the~user commands}
When sending a~user command, the~user fills the~structure containing the~code of the~command, its argument, the~number of
retries and the~timeout for each retry.
Those parameters allow the~user to adjust the~behavior of the~driver to the~expected time of execution of the~command.
The~pointer to this structure is used as the~second argument to the~\wzcode{ioctl} call.
When the~ioctl \wzcode{L3\_V1\_IOC\_USERCMD} is executed, the~current thread is put to sleep until the~command response
is received or until the~timeout expires. In the~latter case, the~command is resent until the~given number of retries is
reached. Together with the~functionalities of the~FPGA core described in subsection~\ref{sec:fpga-command-execution},
this implementation ensures correct single execution of the~command, 
even if either the~command packet or response packet gets lost.

\section{Tests and results}

The~FADE-10G protocol was tested in different scenarios.
The~operation at 1~Gbps was verified using an~Atlys board~\wzcite{url-atlys} and Dell Vostro 3750 (Intel Core i7-2630QM CPU with 2.0~GHz clock).
10~Gbps operation was verified using the~KC705 board~\wzcite{url-kc705} and a~computer
equipped with an~Intel Core i5-4440 CPU with 3.10~GHz clock.
Operation with four 10~Gbps links was verified with an~AFCK board~\wzcite{url-afck} equipped with an FMC board with 4 SFP+ cages.
The board was connected to a~computer equipped with an Intel Xeon CPU E5-2630 v2 with 2.60~GHz clock.

Correctness of transmission was tested with the~FPGA core sending a~preprogrammed sequence of data 
that was later verified by the~receiving computer. Transmissions up to 10~Tb were tested, and no
transmission errors occurred.
The~achievable transmission speed was equal to 990.34~Mbps with 1~Gbps interface in the~Atlys board.

In tests of maximum transmission speed with 10~Gbps links, it was found that verification of data 
led to a decrease of the~achievable throughput. The~user application was not able to process 
data at the~full speed, and the~congestion avoidance algorithm was activated. Therefore, the~maximum 
throughput tests were performed without a~full data verification.
Tests with the~10~Gbps interface in the~KC705 board demonstrated a~throughput of 9.815~Gbps. However to
achieve such a~throughput it was necessary to decrease receive interrupt latency in the~network adapter 
with the~``\wzcode{-C~rx-usecs~0}'' command of the~\wzcode{ethtool} program. With a~standard interrupt latency 
of 1~$\mu$s, the~achievable throughput was equal to only 6.5~Gbps.

The~CPU load was measured with the~\wzcode{top} program. The load during the~transmission via a~single link (without data verification)
was measured on the~computer with an~Intel Core i5 CPU. The result was equal to 2.15\% (0.97\% in user processes, 0.31\% in system processes and 0.87\% in software interrupts).

Operation with four 10~Gbps links in the~AFCK board working simultaneously has shown limitations related
to the~computer speed.
The~achieved mean throughput was equal to 9.72~Gbps per link. With full data verification, the~throughput
was further limited to 8.88~Gbps per link.
The~CPU load measured during the~transmission through four links without verification was equal to 3.90\% (0.10\% in system processes and 3.80\% in software interrupts; load in user processes was reported as 0\%).
The measurements were performed on the~computer with an~Intel Xeon CPU.

In the~final measurement system, the~data delivered to the~memory mapped buffer should be split into records
to be routed to computers in the~DAQ network. With the~appropriate organization of the~readout format, the~separation
of records should not require checking of each received word. 
Therefore, it is expected that the~CPU load related to the~routing of data should be significantly lower than the~load
generated by the~full data verification.
Furthermore, transmission of records to the~DAQ network via modern NICs should be done using the
DMA with minimal CPU load. 
However, implementation of the~readout protocol and the~data routing application is beyond the~scope of this paper as a~responsibility of the~end user.

In summary, the~presented test results suggest, that the~FADE-10G system should allow reception and routing of one
10 Gbps data stream with a~computer with an~Intel Core i5 CPU, and 4 such streams with a~computer with an~Intel Xeon CPU.

Of course, the~CPU computational power is not the~only bandwidth limiting factor. When designing the~system, it is 
necessary to consider throughput of all I/O interfaces involved in the~transfer of data.
To check the~behavior of the~FADE-10G system in conditions where the~output channel limits the~overall bandwidth,
another set of tests was performed with the~user application writing the~data directly to the~SATA SSD disk.
In this setup, the~throughput was limited by the~disk.
The throughput achieved was the~same as for the~application
 storing the~pre-generated pseudorandom data to the~disk. 
The~congestion avoidance algorithm has correctly limited
the~transmission rate from the~FPGA core. The~data stored on the~disk was later analyzed, and no corrupted
data were discovered.

The~last series of tests verified transmission of the~user commands in the~worst conditions.
The~dedicated application transmitted user commands during reception of the~continuous stream of data.
In the~1~Gbps setup, the~FPGA core
was able to execute 2870 commands per second with data transmission speed unaffected.
In the~10~Gbps setup, the~FPGA core was able to execute ca. 40000 commands per second without impairing 
the~transmission speed.

The~protocol has been optimized for operation at high data rate. For lower data rates, there is a~danger that
data may wait too long until the~data packet is completed and transmitted. To avoid this,
the~data source should provide a minimum flow of data (e.g. ``time stamps'' or ``dummy data'')  in the~low data rate conditions. That ensures that the~delay, associated with the~completion
of the~data packet, is acceptable. This approach keeps the~protocol as simple as possible but requires
the~user to enforce the~minimal required data flow in the~upper layer. As that additional data is inserted
only in low traffic conditions, its removal in the~receiver application should not significantly
increase the~CPU load. The~alternative solution could be sending incomplete packets after a~user defined timeout.
Such a~packet could have a~similar payload as the~``last packet'', shown in Table~\ref{tab:fade-frame-payload}d, but should be marked as another packet type (e.g. \wzhex{0xa5a7}).
Unfortunately, handling such packets at the~receiver side would break the~idea of an~efficient,
transparent passing of the received data stream into the~memory buffer directly available for the~user application and would result in a~drop of efficiency at a~high data rate.

\section{Conclusions}
  The~presented FADE-10G system allows reliable transmission of measurement data from an FPGA
   via a~1~Gbps or 10~Gbps interface
  to a~computer running Linux OS. The~system can almost fully utilize the~link throughput.
  Apart from transmission of data, the~system implements simple control or diagnostic commands, which are reliably
  transmitted to the~FPGA. The results of the command execution are also reliably transferred to the~computer.
  Even with a~fully occupied link the~system executes over 2800 commands per second with 1~Gbps link, and ca.
  40000 commands per second with 10~Gbps link.
  The~system minimizes packet acknowledgment latency that in turn allows the~reduction of the~amount of memory needed in the~FPGA
  to buffer the~data. Additionally, the~system implements a~special ``early retransmission'' mechanism, which 
  reduces the~latency of the~data retransmission in case of a~lost packet.
  The~data received by the~computer is delivered to the~user application using the~memory mapped
  kernel buffer, that avoids unnecessary data copying and reduces the~CPU load.
 
  The~FADE-10G system is implemented in a~possibly simple way and published under permissive licenses 
  (most parts under BSD license, some under GPL license and some as public domain). Therefore, it can be
  a~good base solution for further development of a~transmission system suited to a~particular experiment.
  Sources of the~FADE-10G project are available on the~OpenCores website~\wzcite{url-opencores-fade}.
\acknowledgments
 The author thanks Dr Grzegorz Kasprowicz and Dr Dawid Rosołowski from Warsaw University of Technology for 
 providing access to the~hardware needed to test the~FADE-10G system with 10~Gbps Ethernet links.
\bibliography{fade10g}   
\bibliographystyle{JHEP}
\end{document}